\documentclass[compressed,submission]{ieee}

\title{Polar Polytopes and Recovery of Sparse Representations}

\author{Mark D. Plumbley\member{Member}%
\thanks{Manuscript created 13 October 2005}%
\authorinfo{Department of Electronic Engineering,
	Queen Mary University of London,
	Mile End Road, London E1 4NS, UK.
	Email: mark.plumbley@elec.qmul.ac.uk}}

\usepackage{amssymb,amsmath,amsfonts}
\usepackage{graphicx}
\usepackage{url}

\renewcommand{\a}{\vec a}							%
\renewcommand{\c}{\vec c}							%
\renewcommand{\r}{\vec r}					%
\renewcommand{\v}{\vec v}							%
\renewcommand{\vec}[1]{{\mathbf #1}}	%

\newcommand{\mat}[1]{{\mathbf #1}}		%
\newcommand{\setfont}[1]{{\cal #1}}		%

\newcommand{\reals}{{\mathbb R}} %

\newcommand{\setI}{\setfont{I}}		%

\DeclareMathOperator{\aff}{aff}		%
\DeclareMathOperator{\conv}{conv}		%
\DeclareMathOperator{\diag}{diag}		%
\DeclareMathOperator{\mean}{mean}		%
\DeclareMathOperator{\sign}{sign}		%
\DeclareMathOperator{\Spark}{Spark}		%
\DeclareMathOperator{\relint}{relint}		%
\newcommand{\pinv}[1]{#1^\dag}

\newcommand{\norm}[2][]{{\|#2\|}_{#1}}
\newcommand{\Norm}[2][]{{\left\|#2\right\|}_{#1}}
\newcommand{\innerprod}[2]{\langle #1,#2\rangle}
\newcommand{\floor}[1]{\lfloor #1\rfloor}
\newcommand{\chs}[2]{\left(\begin{smallmatrix}#1\\#2\end{smallmatrix}\right)}

\newcommand{\vecone}{\vec 1}		%
\newcommand{\veczero}{\vec 0}		%
\newcommand{\atilde}{\tilde\a}					%
\newcommand{\A}{\mat A}					%
\newcommand{\Aopt}{\A_{\mathrm{opt}}}	%
\newcommand{\Atilde}{\tilde\A}	%
\newcommand{\Atildeopt}{\Atilde_{\mathrm{opt}}}	%
\newcommand{\copt}{\vec c_{\mathrm{opt}}}					%
\newcommand{\cF}{\vec c_{\mathrm{F}}}					%
\newcommand{\Fstaropt}{F^*_{\mathrm{opt}}}	%
\newcommand{\Lone}{\ell_1}
\newcommand{\Lzero}{\ell_0}
\renewcommand{\Phi}{\undefined}	%
\renewcommand{\Psi}{\undefined}	%
\renewcommand{\phi}{\undefined}	%
\renewcommand{\psi}{\undefined}	%
\newcommand{\vecsigma}{\vec\sigma}					%
\newcommand{\x}{\vec x}					%
\newcommand{\xzero}{\x_0}					%
\newcommand{\xtilde}{\tilde\x}	%
\newcommand{\xtildei}{\tilde x_i}	%
\newcommand{\xtildej}{\tilde x_j}	%
\newcommand{\xtildeopt}{\xtilde_{\mathrm{opt}}}	%
\newcommand{\xopt}{\x_{\mathrm{opt}}}	%
\newcommand{\y}{\vec y}					%
\newcommand{\yhat}{\hat\y}			%

\newcommand{\pinva}{\pinv\a}						%
\newcommand{\pinvAopt}{{\pinv\Aopt}}	%
\newcommand{\pinvAtildeopt}{{\pinv{\Atildeopt}}}	%

\newcommand{\intersect}{\cap}

\newtheorem{theorem}{Theorem}[section]		%
\newtheorem{lemma}[theorem]{Lemma}			%
\newtheorem{corollary}[theorem]{Corollary}			%

\newcommand{\ie}{i.e.\ }
\newcommand{\eg}{e.g.\ }

\begin{document}

\maketitle

\begin{abstract}
Suppose we have a signal $\y$ which we wish to represent
using a linear combination of a number of basis atoms $\a_i$,
$\y=\sum_i x_i \a_i = \A\x$.
The problem of finding the minimum $\Lzero$ norm representation
for $\y$ is a hard problem.
The Basis Pursuit (BP) approach proposes to find the minimum
$\Lone$ norm representation instead,
which corresponds to a linear program (LP)
that can be solved using modern LP techniques,
and several recent authors have given conditions for
the BP (minimum $\Lone$ norm) and sparse (minimum $\Lzero$ solutions)
representations to be identical.
In this paper, we explore this \emph{sparse representation problem}
using the geometry of convex polytopes,
as recently introduced into the field by Donoho.
By considering the dual LP
we find that the so-called polar polytope $P^*$
of the centrally-symmetric polytope $P$ 
whose vertices are the atom pairs $\pm \a_i$ is particularly helpful
in providing us with geometrical insight
into optimality conditions given by Fuchs and Tropp
for non-unit-norm atom sets.
In exploring this geometry we are able to tighten some of these earlier results,
showing for example that the Fuchs condition is both necessary and sufficient
for $\Lone$-unique-optimality,
and that there are situations where Orthogonal Matching Pursuit (OMP)
can eventually find all $\Lone$-unique-optimal solutions with $m$ nonzeros even if ERC fails
for $m$, if allowed to run for more than $m$ steps.
\end{abstract}
\begin{keywords}
Sparse representations, Basis Pursuit (BP), Orthogonal Matching Pursuit (OMP),
linear programming, polytopes.
\end{keywords}

\section{Introduction}

Suppose we have a vector $\y = [y_1,\ldots,y_d]^T$ which we wish to represent
using a linear combination from $n$ nonzero $d$-dimensional basis atoms $\a_i$,
$\y=\sum_i x_i \a_i$.
In other words, 
we wish to find an $n$-vector $\x=[x_1,\ldots,x_n]^T$ such that $\y=\A\x$,
where $\A=[\a_i]$ is the $d\times n$ matrix whose $i$th column is $\a_i$.
Unless specified otherwise, the vectors $\a_i$ are not required to be unit norm,
i.e. $\norm[2]{\a_i} \ne 1$ in general.
In the special case where the $\a_i$ are unit norm, 
we may call $\A$ a \emph{dictionary} \cite{MallatZhang93-matching}.

We consider the case where we
have more atoms $\a_i\in\A$
than observation dimensions, $n>d$,
and there are therefore many possible representations $\A\x=\y$ for a given $\A$ and $\y$.
The {\em sparse representation problem} is then to find
the representation $\x$ with the fewest possible
non-zero components,
\begin{equation}
	\min_\x \norm[0]{\x}	\quad\text{such that}\quad \A\x = \y 			\tag{P0}\label{eqn:P0}
\end{equation}
where $\norm[0]{\x}$ is the $\Lzero$ norm of $\x$,
\ie the number of non-zero elements.
This is well known to be a hard problem \cite{ChenDonohoSaunders98-atomic}.

In the signal processing community,
Chen, Donoho and Saunders \cite{ChenDonohoSaunders98-atomic}
proposed to approximate \eqref{eqn:P0} with the `relaxed' $\Lone$ problem
\begin{equation}
	\min_\x \norm[1]{\x}	\quad\text{such that}\quad \A\x = \y 			\tag{P1}\label{eqn:P1}
\end{equation}
where $\norm[1]{\x} = \sum_i |x_i|$ is the $\Lone$ norm of $\x$.
Problem \eqref{eqn:P1},
which they called \emph{Basis Pursuit} (BP),
can be formulated as a linear programming (LP) problem,
which can be solved using well known optimization methods such as the simplex method
or interior point methods 
\cite{Wright04-interior-point}.
They observed experimentally that the solution to \eqref{eqn:P1}
often found a `good' sparse representation for $\y$,
and gave examples where it produced better results than
the greedy algorithms \emph{Matching Pursuit} (MP) \cite{MallatZhang93-matching}
or \emph{Orthogonal Matching Pursuit} (OMP) \cite{PatiRezaiifarKrishnaprasad93-omp}.

Subsequently a number of authors have explored the conditions under which
the minimum of \eqref{eqn:P1} is unique and identical to the minimum of
\eqref{eqn:P0}, 
sometimes called \emph{exact recovery} or \emph{$\Lone/\Lzero$ equivalence}.
For example, 
for dictionaries of unit norm atoms,
Donoho and Huo \cite{DonohoHuo01-uncertainty}
showed that 
if $\A$ is the union of a pair of orthonormal `time' and `frequency' (spike and Fourier)
bases, so that $n=2d$,
then 
$\Lone/\Lzero$ equivalence holds
for a representation $\y=\A\x$
if $\x$ has 
$m = \norm[0]{\x} < \frac{1}{2}\sqrt{d}$ nonzeros.
With $M \triangleq \max_{i\ne j} |\innerprod{\a_i}{\a_k}|$
defined to be the \emph{coherence} of the dictionary,
they also showed that $\Lone/\Lzero$ equivalence holds if
$m < \frac{1}{2}(1+M^{-1})$ \cite{DonohoHuo01-uncertainty}.
Elad and Bruckstein \cite{EladBruckstein02-generalized}
improved this bound to 
$m < (\sqrt{2} - 0.5)M^{-1} = 0.9142 M^{-1}$
for a pair of orthonormal bases,
and
Donoho and Elad \cite{DonohoElad03-optimally}
and independently Gribonval and Nielsen \cite{GribonvalNielsen03-sparse-representations}
generalized these bounds
for more general dictionaries of non-orthogonal unit-norm vectors.

\subsection{Recovery conditions on general, non-unit-norm atom sets}

In this paper we will consider the more general case of non-unit-norm atom sets.
In the longer term we are interested in learning appropriate atom sets and may not
want to constrain these to be unit norm.
Also, for the purposes of the current paper, 
the usual unit-norm requirement on atoms means that the $d=2$ case is somewhat
`too well behaved', making construction of simple 2D visualizations more difficult than necessary.

For clarity it can be helpful to decompose $\Lone/\Lzero$ equivalence
for a representation $\xzero$ into two separate conditions:
\begin{enumerate}
	\item $\xzero$ is the unique optimum to \eqref{eqn:P0}	($\Lzero$-unique-optimality)	\label{itm:L0}
	\item $\xzero$ is the unique optimum to \eqref{eqn:P1}	($\Lone$-unique-optimality)		\label{itm:L1}
\end{enumerate}
To show $\Lone/\Lzero$ equivalence for a given $\xzero$ it is sufficient to show both
$\xzero$ satisfies both $\Lzero$-unique-optimality
and $\Lone$-unique-optimality.
To show $\Lone/\Lzero$ equivalence for a set of representations,
it is sufficient to show both conditions hold for all representations $\xzero$ in that set.
Let us deal with $\Lzero$-unique-optimality first. 

We define the \emph{Spark} of a matrix, $\sigma = \Spark(\A)$,
to be the smallest number such that 
there exists a subset of $\sigma$ columns from $\A$ that are linearly dependent \cite{DonohoElad03-optimally}.
Given a matrix $\A\in \reals^{d\times n}$ with $n>d$,
if all subsets of $d$ columns from $\A$ are linearly independent,
then $\Spark(\A)=d+1$.

\begin{theorem}[Donoho and Elad \protect{\cite[Corollary 1]{DonohoElad03-optimally}}: $\Lzero$-Uniqueness]
\label{thm:L0-unique}
A representation $\y=\A\xzero$ 
with $m = \norm[0]{\xzero}$ nonzeros
is $\Lzero$-unique-optimal (\ie the sparsest possible representation)
if $m<\Spark(\A)/2$.
\end{theorem}

In particular this means that if all subsets of $d$ columns of $\A\in \reals^{d\times n}$ 
are linearly independent, then Theorem \ref{thm:L0-unique} holds
with $m<(d+1)/2$.
Consequently the combination of
$\Lone$-unique-optimality and $m<\Spark(\A)/2$ is sufficient to show $\Lone/\Lzero$ equivalence.
In the remainder of this paper we will therefore concentrate on the condition of 
$\Lone$-unique-optimality.

For the case of general (non-unit-norm) sets of real atoms, 
authors including Tropp \cite{Tropp04-greed} and Fuchs \cite{Fuchs04-sparse}
have derived conditions for $\Lone$-unique-optimality.
We begin with the condition introduced recently by Fuchs \cite{Fuchs04-sparse}.
For some $\y$ represented by a linear combination of $m<d$ atoms in $\A$, let
$\xzero$ be the desired solution of $\y=\A\xzero$ to be recovered,
with $m = \norm[0]{\xzero}$ non-zero elements.
\begin{theorem}[Fuchs Condition \protect{\cite[Theorem 4]{Fuchs04-sparse}}] 
\label{thm:Fuchs}
Let $\xopt$ be the $m$-dimensional vector built from the nonzero components of $\xzero$,
with $\Aopt$ the $n \times m$ matrix built from the corresponding columns of $\A$
such that
$\y = \Aopt \xopt = \A\xzero$.
If $\Aopt$ is full rank,
and there exists some $\c\in\reals^d$ satisfying 
\begin{align}
	\Aopt^T\c &= \sign{\xopt}	\\
	|\a_j^T \c| &< 1 \qquad\text{for any $\a_j\in\A$, $\a_j\notin\Aopt$}	\label{eqn:Fuchs}
\end{align}
then $\xzero$ is the unique solution to \eqref{eqn:P1}.
\end{theorem}

This means that if $\y=\A\xzero$ is a sparse representation of $\y$
such that the conditions in Theorem \ref{thm:Fuchs} hold,
then Basis Pursuit (BP) will find this sparse representation.
For an extension of Theorem \ref{thm:Fuchs} to the complex domain see Tropp \cite{Tropp05-recovery}.

Using $\c = \pinvAopt^T\sign{\xopt}$ in Theorem \ref{thm:Fuchs},
where $\pinvAopt$ is the Moore-Penrose pseudoinverse of $\Aopt$,
we obtain the following result (introduced originally in \cite{Fuchs98-detection}):
\begin{corollary}[Fuchs Corollary \protect{\cite{Fuchs98-detection}}]
\label{cor:FuchsCor}
Let $\xopt$ and $\Aopt$ be given as in Theorem \ref{thm:Fuchs}.
If $\Aopt$ is full rank,
and
\begin{equation}
	|\a_j^T \pinvAopt^T\sign{\xopt}| < 1 \qquad\text{for any $\a_j\in\A$, $\a_j\notin\Aopt$}	\label{eqn:FuchsCor}
\end{equation}
then $\xzero$ is the unique solution to \eqref{eqn:P1}.
\end{corollary}

The conditions involved in Theorem \ref{thm:Fuchs} and Corollary \ref{cor:FuchsCor}
seem at first somewhat awkward to visualize,
in that they involve the sign of $\xopt$ as well as its support
\cite{GribonvalNielsen03-spie,Tropp05-recovery}.
However, we shall show in this paper that they corresponds to finding points $\c$
on a particular geometrical object, the {\em polar polytope},
whose vertices and faces correspond to signed support basis sets.
We shall also show that the condition in Theorem \ref{thm:Fuchs} is the weakest possible,
in that it is both necessary and sufficient for $\Lone$-unique-optimality.

Perhaps more well known than the Fuchs condition above is
the \emph{Exact Recovery Condition} (ERC) 
introduced by Tropp \cite{Tropp04-greed}.
\begin{theorem}[Tropp \protect{\cite{Tropp04-greed}}: Exact Recovery Condition]
\label{thm:erc}
Let us have $\xzero$ and $\Aopt$ as in Theorem \ref{thm:Fuchs} above.
If
\begin{equation}
	\max_{\a_j \notin \Aopt} \Norm[1]{ \pinvAopt \a_j } < 1		\label{eqn:erc}
\end{equation}
where $\a_j$ ranges over the atoms in $\A$ which are not in the $m$-term representation of $\y$,
then $\xzero$ is the unique solution to \eqref{eqn:P1}.
\end{theorem}
Hence a representation $\y=\Aopt\xopt$ can be recovered by BP whenever \eqref{eqn:erc} is satisfied.
The quantity $\max_{\a_j \notin \Aopt} \Norm[1]{ \pinvAopt \a_j }$ is referred to as the
\emph{exact recovery coefficient}.

Tropp \cite{Tropp04-greed}
also showed that \eqref{eqn:erc} guarantees that
the Orthogonal Matching Pursuit (OMP) algorithm
will find the solution $\xzero$ in $m$ steps.
This condition also applies for exponential convergence 
of ordinary matching pursuit (MP) to the solution $\xzero$
\cite{GribonvalVandergheynst04-exponential}.

Although the approaches of Fuchs \cite{Fuchs04-sparse} and Tropp \cite{Tropp04-greed} are very different, Gribonval and Nielsen \cite{GribonvalNielsen03-spie}
pointed that they are closely linked. Specifically we have
\begin{equation}
	\max_{\xopt} \max_{\a_j \notin \Aopt} | \sign(\xopt^T)  \pinvAopt \a_j |
		= \max_{\xopt} \max_{\a_j \notin \Aopt} | \innerprod{\sign(\xopt)}{\pinvAopt \a_j} |	
		= \max_{\a_j \notin \Aopt} \Norm[1]{\pinvAopt \a_j}
		\label{eqn:fuchs-tropp}
\end{equation}
so the Exact Recovery Condition (Theorem \ref{thm:erc})
is itself a corollary of the Fuchs Corollary (Corollary \ref{cor:FuchsCor}).
Thus ERC is a stronger condition than the Fuchs Condition (Theorem \ref{thm:Fuchs}), 
and there are in fact cases where OMP will not give the same solution as BP.

In an interesting new direction,
Donoho \cite{Donoho04-neighborly,Donoho05-high-dimensional} has explored the link between
sparse recovery and the geometry of \emph{polytopes},
convex sets defined by a finite set of vertices or inequalities.
Donoho showed that $\Lone/\Lzero$ equivalence of certain representations $\xzero$
can be linked to the existence of particular faces of a polytope $P$
whose vertices are the atom pairs $\pm\a_i$ with $\a_i\in\A$.
If each subset of $k$ signed atoms forms the vertices of a true face of $P$,
(\ie\ $P$ is $k$-neighbourly)
then $\Lone/\Lzero$ equivalence holds for
all representations $\xzero$ with at most $k$ nonzeros.

This powerful new approach means that results from the field of polytopes can be
brought across to the sparse representations problem, and vice versa.
For example, using the classic work of 
McMullen and Shephard \cite{McMullenShephard68-diagrams}
on centrally symmetric polytopes,
Donoho showed \cite[Corollary 1.3]{Donoho04-neighborly}
the surprising result
that for $n-2\ge d > 2$, the condition
$k\le \floor{(d+1)/3}$ must hold
for $\Lone/\Lzero$ equivalence of all representations $\xzero$ having at most $k$ nonzeros.

The structure of this paper is as follows.
In section \ref{sec:geom} we introduce some polytope notation
and discuss the polytope approach to the sparse representation problem.
In section \ref{sec:dual} we consider the dual LP problem
and the corresponding polar (dual) polytope and its visualization.
In section \ref{sec:fuchs} we investigate the 
Fuchs Condition and its geometry on the polar polytope,
and link this to the Donoho results on the primal polytope.
In the subsequent sections we apply this approach to the Fuchs Corollary
and the Exact Recovery Condition,
and consider the special geometry of unit norm dictionaries.
Finally we consider Matching Pursuit algorithms,
before drawing our Conclusions.
For ease of visualization we will use real geometry in this paper,
so all our matrices and vectors will be real.

\section{Polytopes and sparse recovery}
\label{sec:geom}

We will develop some low-dimensional examples to illustrate the recovery conditions described above.
We will see that, even in 2 dimensions, we can gain considerable insight into the geometric
meaning of these recovery conditions.
First let us define some terminology
(see \eg \cite{Grunbaum03-convex}).

Recall that a set $S\subset\reals^d$ is \emph{convex} if it contains all line segments
connecting any pair of points in $S$, \ie
$x,y\in S$ implies $tx+(1-t)y\in S$ for all $0\le t \le 1$.
A point $x\in S$ is called an \emph{extreme point} 
if it cannot be represented as a convex combination of two other points in $S$.
The \emph{convex hull}, $\conv X$, of a subset $X\subset\reals^d$
is the smallest convex set containing $X$.
The \emph{affine hull}, $\aff X$, of a set of points $x_i\in X$ is the
set of affine combinations $x = \sum_i \lambda_i x_i$ for reals $\lambda_i\ge 0$.
A set of points is said to be \emph{affinely independent} if none of the points $x_i$
can be represented by an affine combination of the other points.

A \emph{convex polytope} is a bounded subset of $\reals^d$ that is the set of solutions
to a finite system of linear inequalities. 
We normally omit the qualifier \emph{convex}.
For example, given an $d \times n$ matrix $\A$ and a $d$-vector $\y$, 
the set $P = \{\x \mid \A\x \le \y\}$ is a polytope if it is bounded:
the notation $\A\x \le \y$ means $\a_i^T\x \le y_i$ for all $i$,
where $\a_i$ is a row of $\A$
and $y_i$ is the corresponding element of $\y$.
(Without the boundedness condition, $P$ would be a \emph{polyhedron}.)
We refer to a $d$-dimensional polytope as a \emph{$d$-polytope}.
A \emph{simplex} is the simplest type of polytope,
and is the $d$-dimensional convex hull of some $d+1$ affinely independent points:
we can call this a \emph{$d$-simplex}.

A linear inequality $\a^T\x\le b$ is called \emph{valid} for a polytope $P$ if it holds
for all elements of $P$.
A subset $F$ of $P$ is called a \emph{face} of $P$ 
if $F=\emptyset$ or $F=P$ (the \emph{improper faces}), or
$$
	F=P \intersect \{ \x \mid \a^T\x = b \}
$$
for some valid inequality $\a^T\x \le b$ with scalar $b$.
Faces of dimension $0$, 
and $d-1$ are called
\emph{vertices}
and \emph{facets}, respectively.
The vertices are also the extreme points of $P$.
Faces of dimension $k$ are called $k$-faces:
these correspond to subsets where (at least) $d-k$ of the inequalities
$\{\a_i^T\x \le y_i\}$ hold with equality.
Faces of a polytope are themselves polytopes.

There are two different ways to represent a polytope $P$: 
by inequalities or by vertices.
If using inequalities, each inequality defines a 
halfspace $H_i = \{\y \mid \a_i^T\x\le y_i\}$
and $P$ is therefore the intersection of all the relevant halfspaces $P=\intersect_i H_i$.
This is called the \emph{H-representation} for $P$.
Alternatively, we can use the set of vertices $V = \{v_1,\ldots,v_p\}$
so that the polytope $P = \conv\{v_1,\ldots,v_p\}$ is the convex hull
of the set of vertices $V$:
this is called the \emph{V-representation} for $P$.
Converting from H-representation to V-representation is called the
\emph{vertex enumeration problem},
while converting from V-representation to H-representation is called the
\emph{convex hull problem} (or \emph{facet enumeration problem})
\cite{AvisFukuda92-pivoting}.

\begin{figure}[htbp]
	\centering
		\includegraphics[width=0.5\linewidth]{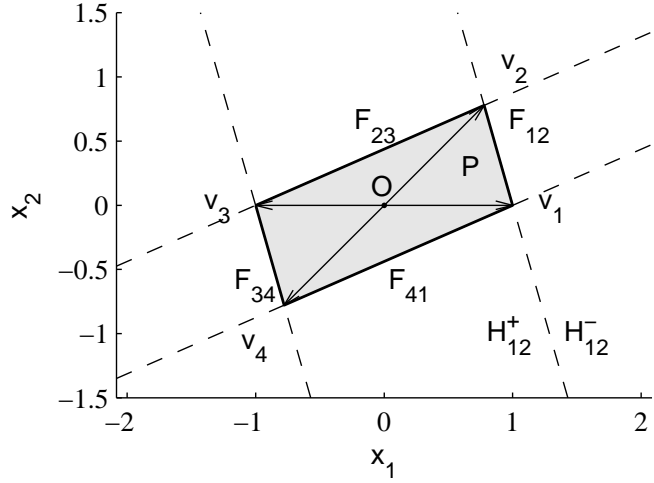}
	\caption{Polytope in two dimensions}
	\label{fig:polytope}
\end{figure}
To summarize some of this terminology, see Fig.~\ref{fig:polytope}.
The $2$-polytope $P$ has been specified
based on its vertices ($0$-faces) $v_1,\ldots,v_4$.
We can visually verify that the polytope $P$ is the convex hull $\conv\{v_1,\ldots,v_4\}$
generated by the vertices of the polytope.
The polytope is also defined by halfspaces such as $H_{12}$: these are shown
as dotted lines, with $H_{12}^+$ indicating the half of $\reals^d$ included in
$H_{12}$, and $H_{12}^-$ the half not included in $H_{12}$.

In fact, Fig.~\ref{fig:polytope} illustrates a specific type of polytope
called a \emph{centrally symmetric} polytope.
A polytope is centrally symmetric if it is symmetric about the origin $O$,
\ie $\x\in P \implies -\x\in P$.
Specifically this means that its vertices come in opposite-sign pairs $(v_i,-v_i)$,
and the inequalities defining the halfspaces also come in opposite-sign pairs.
Thus if the inequality $\a^T\x\le b$ is valid for $P$,
then the negative version $-\a^T\x\le b$ must also be valid.
Centrally-symmetric polytopes are particularly useful
for our consideration of sparse coding.

\subsection{Neighbourliness and sparse recovery}

Now let us form the centrally symmetric polytope $P$ whose $2n$ vertices
are the positive and negative versions of the basis vectors $\pm \a_i$
in our atom matrix $\A$.
We say that the columns of $\A$ are in \emph{general position}
(in this context of defining the vertices of a centrally-symmetric polytope)
if all subsets of $d$ columns of $\A$ are linearly independent
(so $\Spark(\A)=d+1$).

A centrally-symmetric polytope $P$ is called \emph{$k$-neighbourly}
if every subset of $k$ vertices of $P$,
which does not contain two opposite vertices of $P$,
are the vertices of a $(k-1)$-simplex which is a face of $P$.
In other words, for each of the 
$\chs{n}{k} \times 2^k$ 
ways we can choose
a set of $k$ basis vectors $\a_j$ and signs $\sigma_j\in\{-1,+1\}$,
if these $k$ vectors are the vertices of a $(k-1)$-dimensional
face of $P$, then $P$ is $k$-neighbourly.

\begin{theorem}[Donoho \protect{\cite[Theorem 1]{Donoho04-neighborly}}]
Let $P$ be the polytope whose $2n$ vertices are the positive and negative 
atoms $\pm \a_i$ with $a_i\in\A$.
Then $P$ 
is $k$-neighbourly if and only if
every solution $\xzero$ to $\y=\A\xzero$ with at most $k$ nonzeros
is the unique solution to \eqref{eqn:P1}.
\end{theorem}
In other words if $P$ is $k$-neighbourly,
then BP will find all sparse representations $\xzero$ with $\norm[0]{\xzero}\le k$,
i.e. $\xzero$ has at most $k$ nonzero elements.
Results from the theory of convex polytopes \cite{McMullenShephard68-diagrams} then 
give us \eg\ $k\le \floor{(d+1)/3}$ if $n\ge d+2, d>2$.

\begin{figure}[htbp]
	\centering
		(a) \hskip 0.4\linewidth (b) \\
		\includegraphics[width=0.9\linewidth]{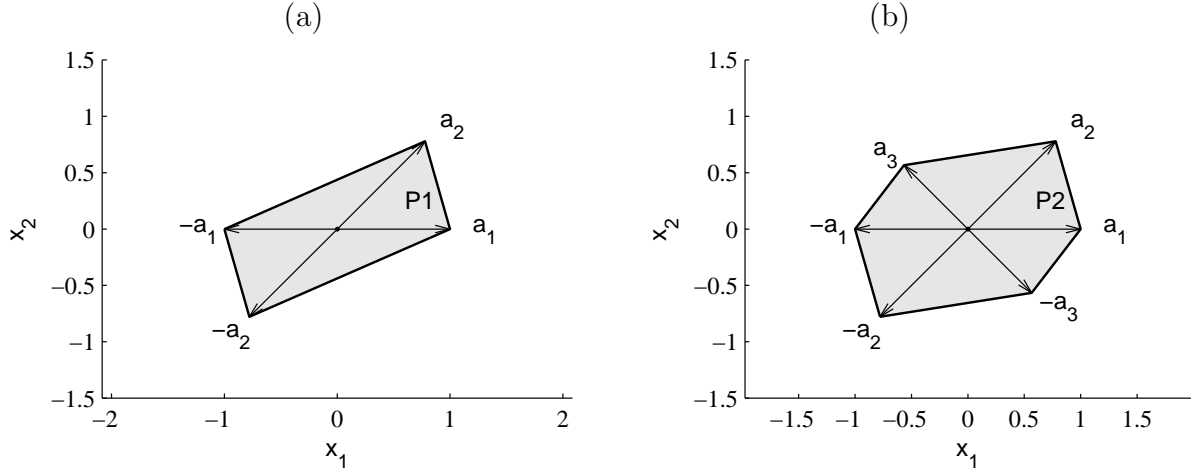}
	\caption{Neighbourly and non-neighbourly polytopes.}
	\label{fig:neighbourly}
\end{figure}
Let us give a visualization of this property in 2 dimensions.
In Fig.~\ref{fig:neighbourly}(a) we have $n=2$ dictionary vectors $\a_1$ and $\a_2$
in $d=2$ dimensions. 
Firstly, we see that the polytope $P1$ has all $2n=4$ vertices.
It is also trivially $k$-neighbourly for $k=1$,
since all 
$\chs{2}{1} \times 2^1 = 4$
ways of choosing a single vertex are the vertices themselves, and hence faces of $P$.
For $k=2$, we can list all
$\chs{2}{2} \times 2^2 = 4$
sets of two vertices (excluding antipodal pairs): these are
$(\a_1,\a_2)$, $(\a_2,-\a_1)$, $(-\a_1,-\a_2)$ and $(-\a_2,\a_1)$.
We can see that each of these 
vertex pairs are the two vertices
of a $1$-face (\emph{edge}) of $P_1$:
the $1$-faces are simply the line segments
between the selected pair of vertices.

In Fig.~\ref{fig:neighbourly}(b) however we have $n=3$ dictionary vectors $\a_1,\a_2,\a_3$,
although still in $d=2$ dimensions.
All $2n=6$ vertices are present, and hence it is again $1$-neighbourly.
However, there are
$\chs{3}{2} \times 2^2 = 12$
ways to choose two vertices, but $P_2$ has only 6 vertices,
so it is not $2$-neighbourly.
For example, 
while the vertex pairs $(\a_1,\a_2)$ and $(\a_1,-\a_3)$ form $1$-faces of $P_2$,
the vertex pairs $(\a_1,\a_3)$ and $(\a_1,-\a_2)$ do not.
Intuitively, we might expect that any $\y$ which is
composed of a positive linear combination of $\a_1$ and $\a_3$
will be unable to be recovered using the linear program \eqref{eqn:P1}.
To gain further insight into this process,
we next introduce a dual polytope
that corresponds to the dual LP of \eqref{eqn:P1}.

\section{Primal-Dual Geometry of Sparse Recovery}
\label{sec:dual}

Authors such as Chen, Donoho and Saunders \cite{ChenDonohoSaunders98-atomic}
and Fuchs \cite{Fuchs04-sparse} have pointed out that
the linear program \eqref{eqn:P1}
has a corresponding dual linear program
\cite{Thie88-introduction,Schrijver98-theory}
\begin{equation}
	\max_\c \c^T\y	\qquad\text{subject to}\qquad \norm[\infty]{\c^T\A} \le 1		\label{eqn:dlp}
\end{equation}
such that for any optimal solution $\xopt$ to \eqref{eqn:P1}
there must be a corresponding optimal solution $\copt$ to \eqref{eqn:dlp}
and this will have the same cost $\copt^T\y = \norm[1]{\xopt}$.
The inequality condition in \eqref{eqn:dlp} can be rewritten
$\norm[\infty]{\c^T\A} \le 1 \equiv |\c^T\a_i|\le 1$ for all $\a_i\in\A$,
or alternatively $+\a_i^T\c\le 1$ and $-\a_i^T\c\le 1$ for all $\a_i\in\A$.
Therefore this dual linear program \eqref{eqn:dlp} defines a second 
polytope $Q=\{\c | +\a_i^T\c\le 1, -\a_i^T\c\le 1 \text{ for all }\a_i\in\A \}$
over the space of $\c$ associated with our dual optimization problem.

To formalize this, we need a little more terminology 
(for details see \eg \cite{Grunbaum03-convex}).
Any polytope $P$ can be associated with a dual polytope $Q$
where each $k$-face of $P$ is associated with a $(d-k-1)$-face of $Q$.
Hence each vertex ($0$-face) of $P$ corresponds to a facet ($(k-1)$-face) of $Q$.
Suppose we have a polytope $P$ with vertices $\v_i\in V$.
The polytope $P^* = \{ \y \mid \v_i^T \y \le 1, \v_i\in V\}$ 
is known as the \emph{polar polytope} of $P$.
If $P$ is a polytope that contains the origin in its interior,
then $P^*$ is also a polytope,
and $(P^*)^* = P$.
Furthermore, vertices, facets, and general $k$-faces of $P^*$
are in a one-to-one correspondence with the facets, vertices, and $(n-k-1)$-faces of $P$, respectively.

Hence the dual polytope specifying the feasible region for $\c$ in \eqref{eqn:dlp}
is simply the polar polytope $P^*$ of our original polytope $P$
whose vertices are the basis vector pairs $\pm\a_i$ with $\a_i\in\A$.
\begin{figure}[htbp]
	\centering
		(a) \hskip 0.4\linewidth (b) \\
		\includegraphics[width=0.9\linewidth]{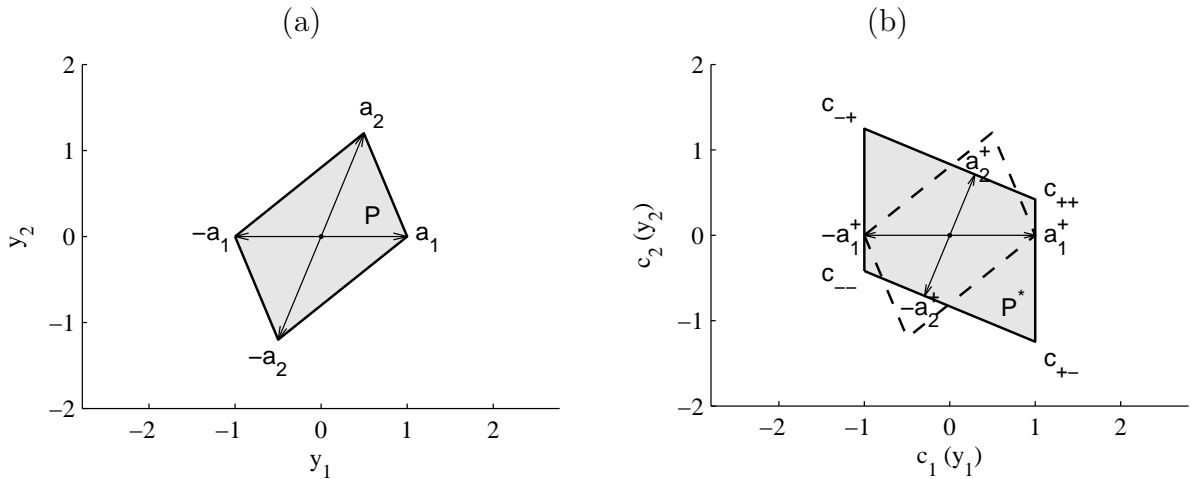}
	\caption{Primal (a) and polar dual (b) polytopes 
	corresponding to the atom set $A=\{\pm \a_1, \pm \a_2\}$}
	\label{fig:primaldual}
\end{figure}
Fig.~\ref{fig:primaldual} illustrates this for the set of basis vectors
$\pm\A = \{\a_1, \a_2, -\a_1, -\a_2\}$.
The facets of the polar (dual) polytope (Fig.~\ref{fig:primaldual}(b))
are along the hyperplanes $\a_i^T\y = 1$.
The vectors $\pm\pinva_i$ shown on the polar polytope figure
are scaled versions of the atoms defined by $\pm\pinva_i = \pm\a_i/\norm[2]{\a_i}^2$.
We notice that the $\pinva_i$ touch the supporting hyperplanes of the dual polytope $P^*$
since $\a_i^T\pinva_i=\a_i^T\a_i/\norm[2]{\a_i}^2 = 1$,
and that $\pinva_i$ is the (transpose of the) Moore-Penrose pseudo-inverse of $\a_i$.
In this particular example we have chosen a unit length atom for $\a_1$,
$|\a_1|=1$, so that $\pinva_1 = \a_1$.

We can also construct a polar polytope for a subset of atoms,
although we have to be slightly careful in this case.
If we choose $m<d$ atoms to generate our primal polytope,
it only occupies at most an $m$-dimensional subspace of $\reals^d$,
and its polar polyhedron (unbounded polytope) extends to infinity.
To avoid this problem we instead introduce 
the concept of a \emph{relative polar polytope} $P^*$ for
an $m$-dimensional polytopes with $m<d$
to be the intersection of the polar polyhedron with the
affine hull of the vertices of $P$
(\ie\ the subspace occupied by the vertices of $P$).
This is therefore the $m$-dimensional polar polytope $P^*$ generated if we considered 
$P$ and $P^*$ both to be restricted to the $m$-dimensional subspace that $P$ occupies.
In what follows, 
where it is clear from context, 
we will simply use `polar polytope' to refer to the relative polar polytope.

\subsection{Primal-dual solution correspondence}

If we have a solution to the dual linear program \eqref{eqn:dlp}
we can find the corresponding solution to the primal linear program \eqref{eqn:P1}
using \emph{complementary slackness}.
To simplify this we will first reformulate \eqref{eqn:P1} and \eqref{eqn:dlp}
into their equivalent standard form.

Let $\xtilde = (\tilde x_1, \dotsc, \tilde x_{2n})^T$ be the nonnegative vector
\begin{equation}
	\xtildei
		= \begin{cases}
				\max(x_i, 0)			&	1 \le i \le n			\\
				\max(-x_{i-n},0)	& n+1 \le i \le 2n
			\end{cases}
				\label{eqn:xtilde}
\end{equation}
and let $\Atilde = [\A,-\A]$ be
the corresponding doubled matrix.
Any solution to $\A\x=\y$ can be written in the form $\Atilde\xtilde=\y$
with nonnegative $\xtilde$.
Using this notation we have $\norm[1]{\x} = \vecone^T\xtilde$ so
we can write the primal and dual problems \eqref{eqn:P1} and \eqref{eqn:dlp} respectively as
\begin{align}
	\min_{\xtilde} \vecone^T\xtilde	&
		\qquad\text{such that}	\qquad \Atilde\xtilde = \y 	\quad\text{and} \quad\xtilde>\veczero	
					\label{eqn:std-P1}	\\
	\max_\c \y^T\c	&
		\qquad\text{such that}  \qquad \Atilde^T\c \le \vecone		\label{eqn:std-dlp}
		. %
\end{align}
Then the complementary slackness of these linear programs gives us 
the following lemma immediately \cite[p95]{Schrijver98-theory}:
\begin{lemma}
Suppose that $\xtilde$ and $\c$ are optimal in \eqref{eqn:std-P1} and \eqref{eqn:std-dlp}.
If a component of $\xtilde$ in \eqref{eqn:std-P1} is positive,
$\xtildei>0$, then we must have equality $\atilde_i^T\c=1$ for the corresponding 
inequality in \eqref{eqn:std-dlp}.
\end{lemma}
Therefore for a given solution $\copt$ we can identify the possible positive
elements of $\xtilde$ by identifying the atoms for which $\atilde_i^T\copt=1$.

\subsection{Brute force algorithm for optimization of \protect{\eqref{eqn:P1}}}

It is a standard result from linear programming that the optimum of the linear function
is obtained at one (or more) of the extreme points \cite{Schrijver98-theory}.
This therefore leads to the following (conceptual) `brute force' algorithm for minimizing the $\Lone$ norm
\eqref{eqn:P1}:
\begin{enumerate}
	\item \label{item:enumv} Enumerate the set $V$ of the vertices of 
		the polar polytope $P^* = \{\c \mid \Atilde^T\c \le \vecone \}$
	\item Search over $V$ to find $\copt = \arg\max_{\c\in V} \c^T\y$.
	\item \label{item:recover} Recover $\Atildeopt$ from $\copt$ 
	and solve for $\xopt = \Atildeopt^{-1}\y$.
\end{enumerate}
We could then recover the basis set $\Atildeopt$ corresponding to $\copt$,
since we have $\atilde_i\notin\Atildeopt$ if $\atilde_i^T\copt<1$
and we consider the remaining rows (for which $\atilde_i^T\copt = 1$)
to be in $\Atildeopt$.
Nonsingular $\Atildeopt$ would indicate non-unique $\xopt$, or no solution.
Alternatively,
if we save the basis sets during vertex enumeration at step \ref{item:enumv},
$\Atildeopt$ can be recovered more directly.
If there were a subspace of optimal solutions for $\c$ which maximize $\c^T\y$
then some of the recovered components of $\xopt$ will in fact be zero.

Now, this algorithm is not meant to be a practical one, particularly since
step \ref{item:enumv} requires the solution of the \emph{vertex enumeration problem}.
The number of vertices of a polynomial can increase very quickly with
the number of facets, 
and the computational and storage complexity of vertex enumeration algorithms
can also be very high \cite{AvisFukuda92-pivoting}.

Nevertheless, it is interesting that this algorithm is very reminiscent of a
clustering algorithm, as if the vertices $\v_i$ of $P^*$ are cluster target vectors,
and we wish to associate $\y$ with the `cluster' (vertex) which `best matches' 
(has largest dot product with) the target.
This may give us a natural way to connect sparse coding with
the \emph{ICA Mixture Model},
which selects between possible representation basis sets depending
on the region occupied by $\y$ \cite{LeeLewickiSejnowski00-ica-mixture}.
Consider also that in many cases the system \eqref{eqn:P1} is to be solved many times for
different observations $\y$.
In this case, it would be possible to `cache' previous known vertices $\c$
and their associated basis sets $\Atildeopt$,
to use as a set of starting points for new solutions.
Observations close to those already found would then be solved immediately,
simply requiring a check for optimality.

\subsection{Visualizing the primal-dual solution correspondence}

\begin{figure}[htbp]
	\centering
		(a) \hskip 0.4\linewidth (b) \\
		\includegraphics[width=0.9\linewidth]{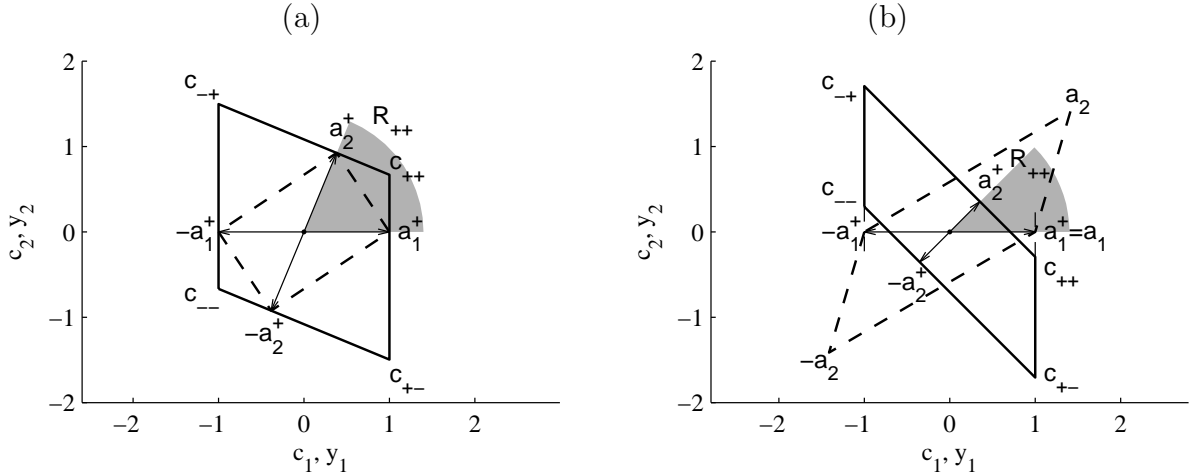}
	\caption{Primal-dual solution correspondence}
	\label{fig:regions}
\end{figure}
Let us consider the subsets of $\{\y \in \reals^d\}$ 
which give particular vertices of $P^*$ and the corresponding
representation basis sets (Fig.~\ref{fig:regions}).
In Fig.~\ref{fig:regions}(a) the shaded region $R_{++}$ denotes a cone in $\y$-space
represented by nonnegative amounts $x_1,x_2\ge 0$ of the basis vectors $+\a_1,+\a_2$.
This segment is bounded by the half-rays in the direction of the corresponding basis
vectors.
It is straightforward to verify that the dot product $\c_{++}^T\y$
of any $\y\in R_{++}$ with the vertex $\c_{++}$
will be larger than the dot product with any other vertex $\c_\setI$,
and hence for any $\y\in R_{++}$, $\c_{++}$ is the point within the polytope
that maximizes $\c^T\y$, as required by the dual linear program (\ref{eqn:dlp})
or its standard form (\ref{eqn:std-dlp}).

Stretching notation slightly,
we may refer to the vertex of the (relative) polar polytope $P^*$
that corresponds to a particular active basis
simply set as the \emph{vertex} of that basis set: 
hence we say that $\c_{++}$ is the vertex of the basis set $\{+\a_1,+\a_2\}$.
In simple cases we find that the vertex $\c_\setI$ is contained 
within the corresponding cone $R_\setI$,
but this is not necessary.
For example in Fig.~\ref{fig:regions}(b)
we see that $\c_{++}$ is not contained in the cone $R_{++}$:
in this case we may say that the basis set has an \emph{external} vertex.

Finally, consider now the observation $\y=\beta (+\a_1)$ for some $\beta>0$,
which has the optimal solution $\x_0 = (\beta, 0)$ corresponding to $\xtilde_0 = (\beta, 0, 0, 0)$.
The quantity $\c^T\y$ is maximized for any $\c$ along the edge joining $\c_{++}$ and $\c_{+-}$, 
\ie\ any $\c \in \conv\{\c_{++},\c_{+-}\}$.
Our brute force algorithm would enumerate the vertices,
so would select either $\copt=\c_{++}$ or $\copt=\c_{+-}$,
and hence determine $\Atildeopt = [+\a_1,+\a_2]$ or $\Atildeopt' = [+\a_1,-\a_2]$ respectively.
But in either case, 
we can confirm that
solving for $\xtildeopt$ would give
$\xtildeopt = \Atildeopt^{-1}\y = (\beta, 0, 0, 0)$
so recovering the desired solution.

\section{The Fuchs Condition}
\label{sec:fuchs}

In its original form, the Fuchs Condition (Theorem \ref{thm:Fuchs}) seems difficult to interpret
(see \eg\ comments in \cite{GribonvalNielsen03-spie,Tropp05-recovery}).
However, if we convert it into its equivalent `standard form'
(in LP terminology) in terms of nonnegative $\xtilde$
then we can relate it more clearly to our polytope geometry.
First however we give the Fuchs Condition in its `standard form',
and show that it is the \emph{weakest possible condition}
for sparse recovery, in that it is both necessary and sufficient
for \eqref{eqn:P1} to find a particular solution to \eqref{eqn:P0}.
In what follows we form $\xtilde$ from $\x$ using \eqref{eqn:xtilde}
together with the corresponding doubled matrix $\Atilde=[\A,\-\A]$.

\begin{theorem}[Fuchs Condition in standard form]
\label{thm:alt-fuchs}
Let $\xtilde_0$ be a solution of $\Atilde\xtilde=\y,\xtilde>\veczero$.
Let $\xtildeopt$ be the $m$-dimensional vector built from the nonzero components of $\xtilde_0$,
with $\Atildeopt$ the $2n \times m$ matrix built from the corresponding columns of $\Atilde$,
such that
$\y = \Atildeopt \xtildeopt = \Atilde\xtilde_0$.
Then $\xtilde_0$ is the unique optimum point of \eqref{eqn:std-P1}
if and only if
$\Atildeopt$ has full rank and there exists some $\c$ such that
\begin{align}
	\atilde_j^T\c = 1	& \qquad \atilde_j \in \Atildeopt	\\
	\atilde_j^T\c < 1	& \qquad \atilde_j \notin \Atildeopt
\end{align}
where $\atilde_j$ ranges over the columns of $\Atilde$.
\end{theorem}
\begin{proof}
For the `if' direction, the set of feasible solutions $\c$ to \eqref{eqn:std-dlp} must satisfy
$\atilde_j^T\c \le 1$ for all $\atilde_j\in\Atilde$.
Complementary slackness states that the following two statements are equivalent \cite{Schrijver98-theory}:
\begin{enumerate}
 \item $\xtilde$ and $\c$ are optimum solutions of \eqref{eqn:std-P1} and \eqref{eqn:std-dlp}
 \item if a component $\xtildei$ of $\xtilde$ is positive, then the corresponding inequality $\atilde_i^T\c \le 1$ is satisfied with equality, \ie\ $\atilde_i^T\c = 1$.
\end{enumerate}
Now the basis vectors $\atilde_j\in\Atildeopt$ are those for which $\xtildei>0$.
Therefore the condition $\atilde_i^T\c = 1$ for $\atilde_j \in \Atildeopt$
is sufficient to specify that $\xtilde_0$ must be an optimum of \eqref{eqn:std-P1}
and $\c$ must be an optimum of \eqref{eqn:std-dlp}.

Complementary slackness also gives us 
that for optimum solutions $\xtilde$ and $\c$,
if an equation $\atilde_i^T\c \le 1$ is satisfied with strict inequality, $\atilde_i^T\c < 1$,
then the corresponding component $\xtildei$ of $\xtilde$ must be zero.
Therefore the condition 
$\atilde_j^T\c < 1$ for $\a_j \notin \Atildeopt$
requires that \emph{any} optimal solution $\xtilde$ to \eqref{eqn:std-P1}
must have zero components $\xtildej=0$ corresponding to $\a_j \notin \Atildeopt$.
Therefore since $\Atildeopt$ is full rank, the optimal solution is unique and is given by
$\xtildeopt=\pinvAtildeopt\y$.

For the converse, suppose first that $\Atildeopt$ does not have full rank.
Then there is a linear subspace of possible solutions for $\xtildeopt'$ satisfying
$\Atildeopt\xtildeopt'=\y$.
Therefore another solution $\xtildeopt'$ would exist
with smaller or identical cost $\vecone^T\xtildeopt'$
so $\xtildeopt$ could not be the unique minimum.
Hence if $\xtildeopt$ is the unique minimum, then $\Atildeopt$ must have full rank.

For the other conditions, we have a feasible solution $\xtilde_0$ to \eqref{eqn:std-P1}
and we know that $\c=\veczero$ is a feasible solution to 
equations \eqref{eqn:dlp} and \eqref{eqn:std-P1} since $\Atilde^T\veczero = \veczero < \vecone$
so both the primal and dual linear programs have a solution.
Since $\xtilde_0$ is an optimum of \eqref{eqn:std-P1}
then there must be at least one optimum solution $\c$ of \eqref{eqn:std-dlp}.
By complementary slackness, for any $i$ with $\xtildei>0$ and hence $\atilde_i\in\Atildeopt$ we must have $\atilde_i^T\c=1$ for any optimum $\c$.
Furthermore, if $\xtilde_0$ is the unique optimum, then there is no optimum
solution with $\xtildei>0$ with corresponding vector $\atilde_i\notin\Atildeopt$
so there must be a solution, say $\c_i$ for which $\atilde_i^T\c_i<1$.
Any convex combination of these optimal solutions $\c_i$ must also be a optimal solution
so let us choose \eg\ $\c' = \mean\{ \c_i | \atilde_i^T\c_i<1 \}$.
Then $\c'$ is an optimum and $\atilde_i^T\c'<1$ for all $\atilde_i\notin\Atildeopt$.
We have therefore constructed a $\c$ which satisfies the required conditions.
\end{proof}

Let us verify that this is equivalent to the Fuchs Condition.

\begin{lemma}
A vector $\c$ satisfies the Fuchs Condition in Theorem \ref{thm:Fuchs}
if and only if it satisfies the condition in Theorem \ref{thm:alt-fuchs}.
\end{lemma}

\begin{proof}
First we note that $\Aopt$ and $\Atildeopt$ contain identical columns
expect for sign changes so the full rank condition on each is equivalent.

For the other conditions in Theorem \ref{thm:Fuchs},
for $\a_j\in\Aopt$, for which $[\xopt]_j \ne 0$, we have $\a_j^T\c = \sign([\xopt]_j)$.
If $[\xopt]_j>0$ 
we get $\a_j^T\c =1$ and $-\a_j^T\c = -1 < 1$ 
so $\atilde_j=\a_j\in\Atildeopt$, $\atilde_{n+j}=-\a_j\notin\Atildeopt$.
Alternatively if $[\xopt]_j<0$
we get $\a_j^T\c =-1 < 1$ and $-\a_j^T\c =1$ so 
$\atilde_j=\a_j\notin\Atildeopt$, $\atilde_{n+j}=-\a_j\in\Atildeopt$.
For $\a_j\notin\Aopt$, we have $|\a_j^T\c| < 1$
so $-1 < \a_j^T\c < 1$, \ie\ $-\a_j^T\c < 1$ and $+\a_j^T\c < 1$,
thus $\atilde_{n+j}^T\c < 1$ and $\atilde_j^T\c < 1$,
so $\atilde_{n+j}\notin\Atildeopt$ and $\atilde_j\notin\Atildeopt$.

Showing the converse is similarly straightforward,
noting that $\a_j^T\c =1$ and $-\a_j^T\c =1$ can never both be satisfied at once.
\end{proof}

From this equivalence we immediately get the following result.

\begin{corollary}
The Fuchs Condition (Theorem \ref{thm:Fuchs}) is both necessary and sufficient
for a given $\x_0$ to be the unique minimum of \eqref{eqn:P1}.
\end{corollary}

Looking at the Fuchs Condition, we see that in standard form (Theorem \ref{thm:alt-fuchs})
it only depends on $\Atildeopt$,
or in original form (Theorem \ref{thm:Fuchs}) on $\Aopt$ and the signs of $\xopt$.
Thus the following follows immediately.

\begin{theorem} \label{thm:discrete}
The condition for $\xtilde_0$ to be the unique minimum of \eqref{eqn:std-P1}
depends only on the support of $\xtilde_0$.
Or equivalently: The condition for $\x_0$ to be the unique minimum of \eqref{eqn:P1}
depends only on the support of $\x_0$ and signs of $\x_0$ on its support.
\end{theorem}
\begin{proof}
The support of $\xtilde_0$ determines $\Atildeopt$ and hence 
both the rank of $\Atildeopt$ and existence of $\c$ in the Fuchs condition
in the standard form (Theorem \ref{thm:alt-fuchs}).
The support and signs of $\x_0$ determines the support of $\xtilde_0$.
\end{proof}

As noted by Donoho \cite{Donoho04-neighborly} this
`discreteness of individual equivalence' has been observed by previous authors
\cite{DonohoHuo01-uncertainty,MalioutovCetinWillsky04-optimal}.
It means for instance that if a particular $\x_0$ is the unique optimal solution
to \eqref{eqn:P1} with $\y=\A\x_0$, then all $\x'$ with
the same support and signs will also be the respective
unique optimal solution to \eqref{eqn:P1} with $\y=\A\x'$.

\subsection{Geometry of the Fuchs condition}

Let us examine a geometrical interpretation of the preceding theorems in terms of
the polar polytope $P^*$ we introduced earlier.

\begin{theorem} \label{thm:dual-face}
Suppose that $\Atildeopt$ has full rank.
The solution $\x_0$ with $m$ nonzeros in Theorem \ref{thm:alt-fuchs},
is the unique optimum point of \eqref{eqn:std-P1}
if and only if the polar $d$-polytope given by $P^* = \{ \c \mid \Atilde^T\c \le \vecone \}$
has a $(d-m)$-dimensional face $\Fstaropt = \{\c\in P^* | \Atildeopt^T\c = \vecone\}$
specified by the $m$ additional equalities
$\Atildeopt^T\c = \vecone$
\end{theorem}
\begin{proof}
For $0 \le m < d$, the conditions in Theorem \ref{thm:alt-fuchs} 
are equivalent to the requirement for
$\c$ to be in the relative interior of the $(d-m)$-face
$\Fstaropt = \{\c\in P^* | \Atildeopt^T\c = \vecone\}$.
Therefore such a $\c$ exists if and only if the face exists and is nondegenerate.
For $m=d$ the conditions are equivalent to $\c$ being exactly the vertex ($0$-face) 
$\c=(\Atildeopt^{-1})^T\vecone$.
\end{proof}

Consequently the Fuchs condition in either its original form (Theorem \ref{thm:Fuchs})
or its standard form (Theorem \ref{thm:alt-fuchs})
corresponds to the existence of the $(d-m)$-dimensional face of $P^*$ 
in Theorem \ref{thm:dual-face}, since for a $\c$ to exist it 
must be in the relative interior of that face (for $m<d$) 
or be the the single vertex point (for $m=d$).

\subsection{Visualizing the Fuchs Condition}
\label{sec:visualizing-fuchs}

Let us return to Fig.~\ref{fig:regions}, with $\x_0=(\beta,0), \beta>0$ in each of
Fig.~\ref{fig:regions}(a) and (b).
In both figures we have 
$\Fstaropt = \conv\{\c_{++},\c_{+-}\}$
which is the line joining $\c_{++}$ to $\c_{+-}$.
Therefore the Fuchs Condition (Theorem \ref{thm:Fuchs} and Theorem \ref{thm:alt-fuchs})
is satisfied by any $\c$ in the relative interior of this line,
$\c \in \relint \Fstaropt$,
\ie\ any point on the line joining $\c_{++}$ to $\c_{+-}$ 
except for the end points $\c_{++}$ and $\c_{+-}$ themselves.

We notice in passing that $\pinva_1 \in\relint \Fstaropt$ in Fig.~\ref{fig:regions}(a)
but not in Fig.~\ref{fig:regions}(b),
so $\c=\pinva_1$ satisfies the Fuchs Condition in the first case but not the second.
We shall see later that this will distinguish the Fuchs Condition from the Fuchs Corollary
(Corollary \ref{cor:FuchsCor}).

\subsection{Relationship to the primal polytope}

Now $P^*$ is the polar (dual) of the primal polytope $P$ with vertices $\pm\a_i$, $\a_i\in\A$.
Therefore the $(d-m)$-face of the polar polytope
$\Fstaropt = \{\c\in P^* | \Atildeopt\c = \vecone\}$,
which we might call the \emph{dual face},
corresponds to the $(m-1)$-face 
$F_{\mathrm{opt}} = P \intersect \conv\{\atilde_j \in \Atildeopt\}$
of the primal polytope $P$,
\ie\ the corresponding \emph{primal face} \cite{Grunbaum03-convex}.
The dual face on $P^*$ exists and is nondegenerate
if and only if the primal face on $P$ exists and is a simplex.
Therefore we have the following result, echoing the individual equivalence results
of Donoho \cite{Donoho04-neighborly}:

\begin{theorem} \label{thm:primal-face}
Let $\xtilde_0$ be a solution of $\Atilde\xtilde=\y,\xtilde>0$,
with $m$ nonzeros, and let $\xtildeopt$ and $\Atildeopt$ be constructed as before.
Then $\xtilde_0$ is the unique optimum point of \eqref{eqn:std-P1}
if and only if
$F_{\mathrm{opt}} = \conv\{\atilde_j \in \Atildeopt\}$
is an $(m-1)$-face of 
$P$.
\end{theorem}

\begin{proof}
This follows immediately from the preceding arguments,
once we note that $\Atildeopt$ has full rank
if and only if $F_{\mathrm{opt}} = \conv\{\atilde_j \in \Atildeopt\}$
has dimension $(m-1)$ and all $\a_j\in\Atildeopt$ are nonzero.
\end{proof}

To summarize, 
for a given solution $\x_0$ to $\A\x=\y$
with $m=\norm[0]{\xzero}$ nonzeros
to be $\Lone$-unique-optimal,
or equivalently for the nonnegative solution $\xtilde_0$ to $\Atilde\xtilde=\y$,
to be $\Lone$-unique-optimal,
we have the following equivalent conditions:
\begin{enumerate}
\item Fuchs Condition in the standard form (Theorem \ref{thm:alt-fuchs})
\item Fuchs Condition in the original form (Theorem \ref{thm:Fuchs})
\item Existence of nondegenerate dual $(d-m)$-face $\Fstaropt$ of $P^*$ (Theorem \ref{thm:dual-face})
\item Existence of primal face $F_{\mathrm{opt}}$ 
	of $P$ which is an $(m-1)$-simplex (Theorem \ref{thm:primal-face})
\end{enumerate}
Furthermore any $\c$ that satisfies the Fuchs Condition
(Theorem \ref{thm:alt-fuchs} or Theorem \ref{thm:Fuchs})
is contained in the relative interior of the dual face $\Fstaropt$ of $P^*$.

To use our approach to 
confirm the main result of Donoho \cite{Donoho04-neighborly},
suppose $P$ is $k$-neighbourly.
Then all representations $\y=\Atildeopt\xtildeopt$ with $m\le k$ nonzeros have 
a face $F_{\mathrm{opt}}$ of the centrally-symmetric primal polytope $P$
which is an $(m-1)$-simplex.
Therefore the Fuchs condition is satisfied for all $\xzero$ with at most $k$
nonzeros, and we have $\Lone$-unique-optimality.
Note that we have not required the assumption of general position of the columns of $\A$:
the requirement of $k$-neighbourliness of the centrally symmetric $P$ is sufficient
to require linear independence of the columns of all optimal submatrices $\Atildeopt$
with at most $m$ columns, which requires $\Spark(\A)>m$.
Finally for $\Lone/\Lzero$-equivalence we simply need to add the stronger
condition $m<\Spark(\A)/2$, so if $P$ is $k$-neighbourly then
we have $\Lone/\Lzero$-equivalence if $m\le \min(k,\Spark(\A)/2-1)$.

\section{Fuchs Corollary}

Let us write down an equivalent of the stronger Fuchs Corollary
(Corollary \ref{cor:FuchsCor}) in the standard form.
\begin{corollary}[Fuchs Corollary in standard form] \label{cor:std-FuchsCor}
For a desired solution $\xtilde_0$ to $\Atilde\xtilde=\y$,
let us construct $\xtildeopt$ and $\Atildeopt$ as before.
If $\Atildeopt$ has full rank and
\begin{equation}
	\a_j^T \copt < 1 \qquad\text{for all}\quad \a_j\in\Atilde, \a_j\notin\Atildeopt
\end{equation}
is satisfied with the specific dual vector $\copt=\pinvAtildeopt{}^T \vecone$,
then $\xtilde_0$ is the unique optimum to \eqref{eqn:std-dlp}.
\end{corollary}
The dual vector $\copt=\pinvAtildeopt{}^T \vecone$
is the \emph{vertex} of our (signed) basis set $\Atildeopt$.

From our geometric viewpoint,
the Fuchs Corollary requires that the dual face
$\Fstaropt = \{\c\in P^* | \Atildeopt\c = \vecone\}$
corresponding to the signed optimal basis $\Atildeopt$
exists (as for the Fuchs Condition),
and additionally that the basis vertex
$\copt=\pinvAtildeopt{}^T \vecone$
is contained in its relative interior, $\copt \in \relint\Fstaropt$.

From a practical point of view,
one advantage of the Fuchs Corollary over the Fuchs Condition
is that it is easier to test.
The probe point $\copt$ can be constructed directly from $\x_0$ and $\A$,
while testing the Fuchs Condition would require
the relevant face of $P^*$ to be found.

\subsection{Visualizing the Fuchs Corollary}

Consider again Fig.~\ref{fig:regions} with $\x_0=(\beta,0), \beta>0$.
Here we have $\Atildeopt = [\a_1]$
and hence 
$\pinvAtildeopt{}^T = [+\pinva_1]$
so our basis vertex is given by
$\copt = \pinvAtildeopt{}^T \vecone = +\pinva_1 \cdot 1 = \pinva_1$.
Since $\Fstaropt = \conv\{\c_{++},\c_{+-}\}$
which is the line segment joining $\c_{++}$ to $\c_{+-}$,
clearly $\copt \in\relint \Fstaropt$ in Fig.~\ref{fig:regions}(a),
but $\copt \notin\relint \Fstaropt$ in Fig.~\ref{fig:regions}(b).
Therefore, while the Fuchs Condition (Theorem \ref{thm:Fuchs} and Theorem \ref{thm:alt-fuchs})
is satisfied for $\x_0=(\beta,0)$ in both Fig.~\ref{fig:regions}(a) and (b),
the Fuchs Corollary (Corollary \ref{cor:FuchsCor}) 
is only satisfied for this $\x_0$ in Fig.~\ref{fig:regions}(a).
This confirms that the Fuchs Corollary is indeed strictly stronger
than the Fuchs Condition (see also \cite{Fuchs04-sparse}).

\section{Exact Recovery Condition}

We saw in the Introduction that the Exact Recovery Condition (Theorem \ref{thm:erc})
of Tropp \cite{Tropp04-greed}
can be derived as a corollary of the Fuchs Corollary (Corollary \ref{cor:FuchsCor}).
To gain geometrical insight, it is helpful for us to state this in the following way:
\begin{lemma}
Suppose we have a desired solution $\xzero$ to $\y=\A\xzero$.
Then the Exact Recovery Condition (Theorem \ref{thm:erc})
is satisfied if the Fuchs Corollary (Corollary \ref{cor:FuchsCor})
is satisfied for all $\xzero'$ with the same support as $\xzero$,
including solutions $\xzero'$ with the same support but different signs.
\end{lemma}
\begin{proof}
This follows from \eqref{eqn:fuchs-tropp} (see \cite{GribonvalNielsen03-spie}).
\end{proof}

From discreteness of the unique minimum condition (Theorem \ref{thm:discrete})
we only have to test a finite number ($2^m$) separate conditions to check all of
the different signs on a support of $m$ nonzeros.
(In fact since those with entirely reversed signs will have identical results,
we only need $2^{m-1}$ tests.)

One way to see this is to explicitly construct the set of
basis vertices $\{\c=\pinvAopt^T\sign\xopt \}$
that will need to be tested.
To do this, let us construct the signs
$\sigma_j\in\{+1,-1\} \text{ for } j=1,\dotsc,m$
and form the sign vector
$\vecsigma = [\sigma_1,\dotsc,\sigma_m]^T \in \{+1,-1\}^m$.
Then the set of basis vertices we need to test is
$V_{\mathrm{opt}}^* = \{ \c=\pinvAopt^T\vecsigma  \}$
which clearly has $2^m$ elements.
ERC (Theorem \ref{thm:erc}) will therefore be satisfied if
\begin{equation}
		\a_j^T\c < 1	\quad \text{for all} \quad	\c\in V_{\mathrm{opt}}^*, \a_j\notin \Aopt
			\label{eqn:erc-v}
\end{equation}
from which it is clear that each of our $2^m$ `tests'
will in fact require $(n-m)$ dot product calculations each.
For dictionaries of unit-norm atoms other measures such as the mutual coherence
$M = \max_{i\ne j}|\a_i^T\a_j|$ 
can give us more practical conditions that guarantee ERC is satisfied,
such as $m<\frac{1}{2}(1+M^{-1})$
\cite{Tropp04-greed,GribonvalNielsen03-spie}.

\subsection{Geometry of the Exact Recovery Condition}

To turn the preceding condition \eqref{eqn:erc-v} for ERC into a geometric visualization,
we can realize that $\c=\pinvAopt^T\vecsigma$ 
is the vector in the span of the columns of $\Aopt$ which satisfies
$\Aopt^T\c=\vecsigma$ \ie\ $\diag(\vecsigma)\Aopt^T\c=\vecone$,
or in other words $\pm_j\a_j^T\c = 1$ for $\a_j\in\Aopt$ and some combination of signs $\pm_j$.
Hence $V_{\mathrm{opt}}^*$ is actually the set of $2^m$ vertices of the
relative polar polytope $P_{\mathrm{opt}}^*$
whose corresponding primal polytope $P_{\mathrm{opt}}$
has the $2m$ vertices
$\pm\a_j$, $\a_j\in\Aopt$.
We call $P_{\mathrm{opt}}$ the \emph{primal basis polytope}
and $P_{\mathrm{opt}}^*$ the \emph{dual basis polytope}.

Consequently ERC is satisfied if and only if 
(a) the dual basis polytope
$P_{\mathrm{opt}}^*$
is contained within the complete polar polytope $P^*$, $P_{\mathrm{opt}}^* \subset P^*$,
and 
(b) $P_{\mathrm{opt}}^*$ does not touch any face of $P^*$
for which $\pm\a_j^T\c = 1$ for some $\a_j\notin\Aopt$
for full rank $\Aopt$.

\subsection{Visualizing the Exact Recovery Condition}

Consider again Fig.~\ref{fig:regions} with $\x_0=(\beta,0), \beta>0$.
Here we have $\Aopt = [\a_1]$
so our primal basis polytope is given by
$P_{\mathrm{opt}} = \conv\{-\a_1, +\a_1\}$.
The relative polar polytope is given by 
$P_{\mathrm{opt}}^* 
	= \{\c\in\aff P_{\mathrm{opt}} | \c^T\a \le 1 \quad\text{for all}\quad \a\in P_{\mathrm{opt}} \}$
where $\aff P_{\mathrm{opt}}$ is the affine hull of $P_{\mathrm{opt}}$.
In this case we get
$P_{\mathrm{opt}}^* = \conv\{-\pinva_1, +\pinva_1\}$
so $P_{\mathrm{opt}}^*$ is the line segment joining $-\pinva_1$ and $+\pinva_1$.
In Fig.~\ref{fig:regions}(a) we can see that $P_{\mathrm{opt}}^* \subset P^*$
and $P_{\mathrm{opt}}^*$ is well away from the faces 
along $+\a_2^T\c = 1$ (joining $\c_{-+}$ to $\c_{++}$) and 
$-\a_2^T\c = 1$ (joining $\c_{--}$ to $\c_{+-}$).
Hence ERC is satisfied in Fig.~\ref{fig:regions}(a).
However, in Fig.~\ref{fig:regions}(b) we can see that $P_{\mathrm{opt}}^* \not\subset P^*$
so ERC is not satisfied.

If we repeat this analysis for some $\x_0$ with $\Aopt=[\a_2]$,
we see that 
$P_{\mathrm{opt}}^* = \conv\{-\pinva_2, +\pinva_2\}$
so $P_{\mathrm{opt}}^* \subset P^*$,
and $P_{\mathrm{opt}}^*$ is away from the other faces,
in both 
Fig.~\ref{fig:regions}(a) and (b), 
and hence ERC is satisfied for both.
Similarly for some $\x_0$ with $\Aopt=[\a_1,\a_2]$,
we now have 
$P_{\mathrm{opt}}^* = P^*$ so clearly
$P_{\mathrm{opt}}^* \subset P^*$,
and there are no $\a_j\notin\Aopt$ to concern ourselves with.
Hence ERC is again satisfied for both Fig.~\ref{fig:regions}(a) and (b).

This illustrates that it is possible for ERC to be satisfied
for all $\x_0$ with $m$ nonzeros (here $m=2$),
but not satisfied for $\x_0$ with $k<m$ nonzeros
(\eg\ $k=1$ and $\x_0=(\beta,0)$ in Fig.~\ref{fig:regions}(b)).
This is in contrast to the Fuchs Condition
where the property of neighbourliness tells us that
if the Fuchs Condition is satisfied for all $\x_0$ with $m$ nonzeros,
then it will be satisfied for any $\x_0$ with $k<m$ nonzeros \cite{Donoho04-neighborly}.

\section{Unit-norm dictionaries}

Many of the equivalence results of previous authors are for dictionaries
of unit norm atoms $|\a_i|=1$.
The fact that $\pinva_i = \a_i$
leads immediately to a number of special properties,
under the assumption that the atoms are distinct:
\begin{enumerate}
	\item Any unit-norm dictionary has all $2n$ vertices;
	\item ERC is satisfied for any 1-term (singleton) representation;
	\item In $d=2$ all basis vertices are internal;
	\item Any centrally symmetric $2$-polytope with 4 vertices is 2-neighbourly.
\end{enumerate}
The simple proofs of these properties are left as an exercise for the reader.
While these can be useful properties, for visualization purposes it means we
have to work harder to find examples illustrating the distinction between ERC
and the Fuchs Condition.
Nevertheless, let us explore what happens with the following basis set
\begin{align}
	\a_1 &= [1,0,0]^T	\\
	\a_2 &= [0,1,0]^T \\
	\a_3 &= (1/\sqrt{3})[1,1,1]^T
\end{align}
to form the matrix $\A = \{\pm \a_i | i = 1,2,3\}$.
Suppose that our desired vector to recover is $\x_0=[1,1,0]^T$
so that $\y=\A\x_0=\a_1+\a_2$.
Therefore the optimal basis set that we would like to recover given $\y$ is
$\Aopt = [\a_1, \a_2]$,
which has vertex $\copt = \pinvAopt^T\vecone = [1, 1, 0]^T$.

\begin{figure}[htbp]
	\centering
		\begin{tabular}{cc}
			(a) & (b) \\
			\includegraphics[width=0.4\linewidth]{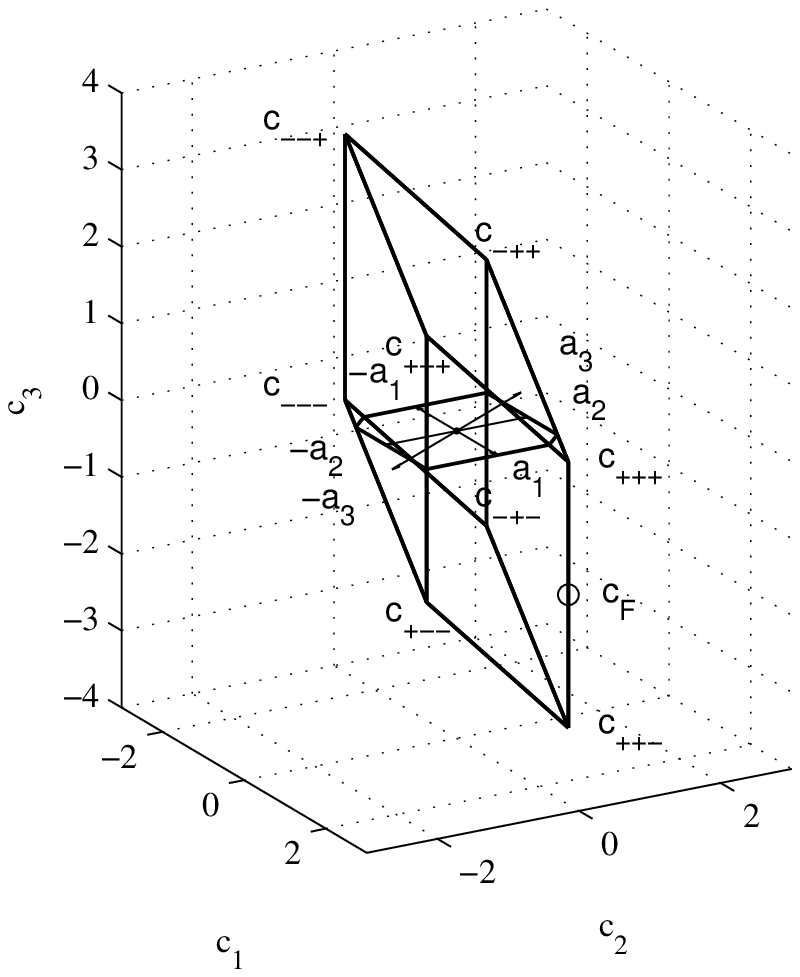}
			&
			\includegraphics[width=0.4\linewidth]{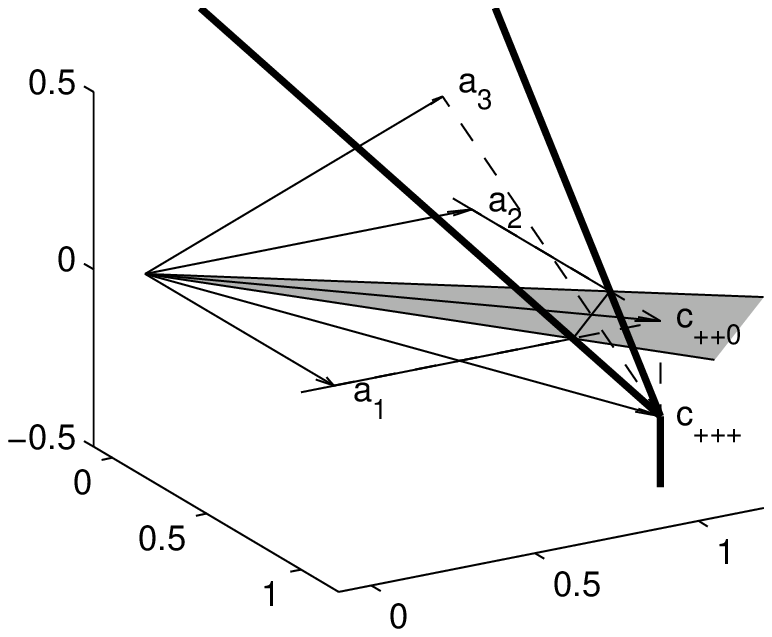}
		\end{tabular}
	\caption{Failure of ERC for unit norm vectors in $d=3$ dimensions,
	showing (a) the complete polar polytope, with section line through the `waist' at $y_3=0$,
	and (b) the magnified section showing the atoms $\a_1$, $\a_2$ and $\a_3$
	($=\pinva_1, \pinva_2$, and $\pinva_3$ since $|a_i|=1$).
	}
	\label{fig:polytope3}
\end{figure}

Consider first the Exact Recovery Condition.
ERC requires that $\norm[1]{\pinvAopt\a_3} = \copt^T \a_3 < 1$
but calculation gives $\copt^T\a_3 = 2/\sqrt{3} > 1$ so ERC fails
for this basis.
We can see this graphically
in Fig.~\ref{fig:polytope3}.
The shaded cone in Fig.~\ref{fig:polytope3}(b)
shows the segment of the plane spanned by $\{\a_1,\a_2\}$
for which $\a_3^T\c > \max_{i=1,2}\a_i^T\c$.
Here we see that the vertex $\copt = \c_{++0}$ 
is in this shaded region (Fig.~\ref{fig:polytope3}(b)), 
and has been `cut off' by the halfspace
$\a_3^T\c \le 1$.

As confirmation of this, the dual basis polytope $P_{\mathrm{opt}}^*$
is the square in the plane $x_3=0$ with vertices at
$[\pm 1, \pm 1, 0]$. 
We can see that
the corner containing $[1,1,0]$ ($=\c_{++0}$)
is not contained within the full dual polytope, 
so $P_{\mathrm{opt}}^* \not\subset P^*$,
and hence ERC is not satisfied.

However, we can identify vectors suitable to satisfy the Fuchs Condition.
For example, consider the point $\cF = [1,1,-2]^T$ marked
in Fig.~\ref{fig:polytope3}(a).
We can verify that $\cF^T\a_1 = \cF^T\a_2 = 1$,
and $|\cF^T\a_3| = |(1+1-2)/\sqrt{3}| = 0 < 1$
therefore the Fuchs Condition is satisfied.
In fact the relevant dual face is
$\Fstaropt = \conv\{\c_{+++},\c_{++-}\}$
so any $\c\in\relint\Fstaropt$,
\ie\ anywhere along the line segment strictly between 
$\c_{+++}$ and $\c_{++-}$,
will be suitable to satisfy the Fuchs Condition.

Finally if we consider the Fuchs Corollary,
this requires $\copt = \pinvAopt^T\vecone = \c_{++0}$ to be 
contained in $\Fstaropt$.
This is clearly not the case, since $\c_{++0} \notin P^*$
and $\Fstaropt$ is itself a face of $P^*$,
so $\Fstaropt\subset P^*$ and therefore $\c_{++0} \notin \Fstaropt$.
Therefore the Fuchs Corollary is not satisfied.

Consequently any desired solution $x_0=[\beta_1, \beta_2, 0]$ with $\beta_1,\beta_2>0$ 
will be recovered by Basis Pursuit, even though ERC and the Fuchs Corollary fails.
Note however that visual inspection of Fig.~\ref{fig:polytope3}(a) will confirm
that both the Fuchs Condition and the Fuchs Corollary would be satisfied for \eg\ 
$x_0=[\beta_1,-\beta_2,0]$ with $\beta_1,\beta_2>0$,
even though ERC must still fail since the support of the desired solution is unchanged.

\section{Matching Pursuit Algorithms}

While we have seen that Tropp's ERC is sufficient but not necessary for 
$\Lone$-unique-optimality,
it really comes into its own for orthogonal matching pursuit (OMP),
as is clear from Tropp \cite{Tropp04-greed}:
\begin{theorem}[Tropp: Exact Recovery for OMP] \label{thm:erc-omp}
Suppose we have a desired solution $\xzero$ for $\y=\A\xzero$
with full rank $\Aopt$ as in Theorem \ref{thm:erc}.
Then Orthogonal Matching Pursuit (OMP) will recover $\xzero$ in $m$ steps
if the Exact Recovery Condition \eqref{eqn:erc} holds.
Conversely, suppose ERC fails for some $\y=\A\xzero$
with optimal synthesis matrix $\Aopt$.
Then there are signals in the column span of $\Aopt$
which Orthogonal Matching Pursuit cannot recover
\emph{in $m$ steps}.
\end{theorem}
\begin{proof}
For the forward direction see \cite{Tropp04-greed}.
For the converse, choose the signal $\y = \copt = (\pinvAopt)^T\vecone$,
for which $\a_j^T\y = 1$ for all $\a_j\in\Aopt$.
If ERC fails there exists some $\a_j \notin \Aopt$ for which
$\a_j^T\y \ge 1 = \max_{\a_i\in\Aopt}\a_i^T\y$.
Therefore OMP may choose this $\a_j\notin\Aopt$ at the first step
(and certainly will if $\a_j^T\y > 1$).
Since we have now used up one step,
and it must take at least $m$ more steps to obtain the correct representation for $\y$,
OMP cannot obtain the correct $m$-term representation in $m$ steps.
\end{proof}

Recovery `in $m$ steps' is implicit in Tropp's statement of this theorem.
However, given that ERC for all desired vectors $\x_0$ with $k$ nonzeros
does not imply ERC holds for all vectors $\x_0$ with $m<k$ nonzeros,
it may still be possible for OMP to recover the $m$-term representation
in some $k>m$ steps,
provided that OMP is eventually allowed to drop any zeros in the final representation.

As an example, consider the situation illustrated in
Fig.~\ref{fig:regions}(b), where we have
$\a_1 = [1, 0]^T$ and $\a_2 = [\sqrt{2}, \sqrt{2}]^T$.
Suppose we wish to recover the signal $\x_0 = [1, 0]^T$ from $\y = \A\x_0 = [1, 0]^T$
for which $\Aopt = [\a_1]$.
Investigating ERC we find
$\pinvAopt \a_2 = \a_1^T\a_2 = \sqrt{2} > 1$
so ERC fails, confirming our earlier discussion.

But let us run OMP to see what happens.
In step 1, OMP chooses the wrong atom $\a_2$, as we now expect,
so $\A^{(1)} = [\a_2]$.
Choosing $x_2$ to minimize the mean squared error we get
$\x^{(1)} = \pinva_2\y = [1/(2\sqrt{2})]$ producing a reconstruction
$\yhat^{(1)} = \x^{(1)} \a_2 = (1/(2\sqrt{2})) \times  [\sqrt{2}, \sqrt{2}]^T = [0.5, 0.5]$
and residual $\r^{(1)} = \y - \yhat^{(1)} = [0.5, -0.5]^T$.
So as expected, OMP has not recovered $x_0 = [1, 0]^T$ in $m=1$ steps.

But if we allow OMP to run for a second step,
we find $\a_1^T\r^{(1)} = 0.5$
while $\a_2^T\r^{(1)} = 0$ as we would expect for OMP.
Hence in step 2, OMP chooses the remaining basis $\a_1$
so $\A^{(2)} = [\a_1, \a_2]$ (reordering the atoms for convenience).
Now choosing $\x = [x_1, x_2]$ to minimize the mean squared error we get
$\x^{(2)} = [x_1^{(2)}, x_2^{(2)}] = (\pinv{\A})^{(2)}\y = [1, 0]^T$ 
producing a reconstruction
$\yhat^{(2)} = \x^{(2)} \A^{(2)} = \y_0$ and $\r = \veczero$.
Since $x_2^{(2)}=0$, OMP has found the correct $1$-term reconstruction of $\y$,
albeit taking 2 steps to do so.

Thus failure of ERC does not require that OMP will fail,
only that it cannot succeed in $m$ steps.
We can therefore state the following weaker condition for \emph{eventual}
recovery by OMP.
\begin{theorem}
Suppose that $\x_0$ with $m_0$ nonzeros is a desired solution of $\y_0=\A\x_0$ which fails ERC.
Suppose further that there exists a different solution $\y_1=\A\x_1$
for which ERC is satisfied, and which \emph{covers} $\x_0$ in the sense that
the support of $\x_1$ is a superset of the support of $\x_0$.
Then OMP will `eventually' recover $\x_0$ in $m_1$ steps, 
where $m_1>m_0$ is the number of nonzeros in $\x_1$
\end{theorem}
\begin{proof}
This follows from the proof of Theorem \ref{thm:erc-omp},
but considering $\x_0$ to be the desired solution within
the extended support given by $\x_1$.
\end{proof}
At present it is unclear whether it is common for ERC to fail at one level $m_0$
but be satisfied at higher levels $m_1>m_0$, so it remains to be seen
whether this concept of \emph{eventual convergence of OMP} will turn out to be useful.

\section{Conclusions}

We have explored the geometry of the sparse representation problem using
centrally-symmetric polytopes and polar (dual) polytopes.
We have seen that polytopes can give us a useful insight into
the optimality conditions introduced by Fuchs, for example,
which had previously been considered to be difficult to interpret.

In exploring this geometry we have also been able to tighten some of these previous results,
and link these to the polytope-based results of Donoho for the primal polytope.
For example, we showed that the Fuchs Condition is both necessary and sufficient
for $\Lone$-unique-optimality,
and that there are situations where Orthogonal Matching Pursuit (OMP) can find
all $\Lone$-unique-optimal solutions with $m$ nonzeros,
even if the Exact Recovery Condition (ERC) fails for $m$,
if it is allowed to run for additional steps.

\section{Acknowledgements}

This work is partially supported by EPSRC grants GR/S82213/01, GR/S75802/01, EP/C005554/1 and EP/D000246/1. Some of the figures were generated using the Multi-Parametric Toolbox (MPT) for Matlab \cite{KvasnicaGriederBaotic04-mpt}.

\bibliographystyle{IEEEtran}
\bibliography{Plumbley05-polar}

\begin{thebibliography}{10}
\providecommand{\url}[1]{#1}
\csname url@rmstyle\endcsname
\providecommand{\newblock}{\relax}
\providecommand{\bibinfo}[2]{#2}
\providecommand\BIBentrySTDinterwordspacing{\spaceskip=0pt\relax}
\providecommand\BIBentryALTinterwordstretchfactor{4}
\providecommand\BIBentryALTinterwordspacing{\spaceskip=\fontdimen2\font plus
\BIBentryALTinterwordstretchfactor\fontdimen3\font minus
  \fontdimen4\font\relax}
\providecommand\BIBforeignlanguage[2]{{%
\expandafter\ifx\csname l@#1\endcsname\relax
\typeout{** WARNING: IEEEtran.bst: No hyphenation pattern has been}%
\typeout{** loaded for the language `#1'. Using the pattern for}%
\typeout{** the default language instead.}%
\else
\language=\csname l@#1\endcsname
\fi
#2}}

\bibitem{MallatZhang93-matching}
S.~Mallat and Z.~Zhang, ``Matching pursuits with time-frequency dictionaries,''
  \emph{IEEE Transactions on Signal Processing}, vol.~41, no.~12, pp.
  3397--3415, 1993.

\bibitem{ChenDonohoSaunders98-atomic}
S.~S. Chen, D.~L. Donoho, and M.~A. Saunders, ``Atomic decomposition by basis
  pursuit,'' \emph{SIAM Journal on Scientific Computing}, vol.~20, no.~1, pp.
  33--61, 1998.

\bibitem{Wright04-interior-point}
M.~H. Wright, ``The interior-point revolution in optimization: History, recent
  developments, and lasting consequences,'' \emph{Bulletin (New Series) of the
  American Mathematical Society}, vol.~42, no.~1, pp. 39--56, 2004.

\bibitem{PatiRezaiifarKrishnaprasad93-omp}
Y.~C. Pati, R.~Rezaiifar, and P.~S. Krishnaprasad, ``Orthogonal matching
  pursuit: Recursive function approximation with applications to wavelet
  decomposition,'' in \emph{Conference Record of The Twenty-Seventh Asilomar
  Conference on Signals, Systems and Computers, Pacific Grove, {CA}}, 1-3 Nov.
  1993, pp. 40--44.

\bibitem{DonohoHuo01-uncertainty}
D.~L. Donoho and X.~Huo, ``Uncertainty principles and ideal atomic
  decomposition,'' \emph{IEEE Transactions on Information Theory}, vol.~47,
  no.~7, pp. 2845--2862, November 2001.

\bibitem{EladBruckstein02-generalized}
M.~Elad and A.~M. Bruckstein, ``A generalized uncertainty principle and sparse
  representation in pairs of bases,'' \emph{IEEE Transactions on Information
  Theory}, vol.~48, no.~9, pp. 2558--2567, September 2002.

\bibitem{DonohoElad03-optimally}
D.~L. Donoho and M.~Elad, ``Optimally sparse representation in general
  (nonorthogonal) dictionaries via \protect{$\ell^1$} minimization,''
  \emph{Proc. Nat. Aca. Sci.}, vol. 100, pp. 2197--2202, March 2003.

\bibitem{GribonvalNielsen03-sparse-representations}
R.~Gribonval and M.~Nielsen, ``Sparse representations in unions of bases,''
  \emph{IEEE Transactions on Information Theory}, vol.~49, no.~12, pp.
  3320--3325, December 2003.

\bibitem{Tropp04-greed}
J.~A. Tropp, ``Greed is good: Algorithmic results for sparse approximation,''
  \emph{IEEE Transactions on Information Theory}, vol.~50, no.~10, pp.
  2231--2242, Oct. 2004.

\bibitem{Fuchs04-sparse}
J.-J. Fuchs, ``On sparse representations in arbitrary redundant bases,''
  \emph{IEEE Transactions on Information Theory}, vol.~50, no.~6, pp.
  1341--1344, 2004.

\bibitem{Tropp05-recovery}
J.~A. Tropp, ``Recovery of short, complex linear combinations via
  \protect{$\ell_1$} minimization,'' \emph{IEEE Transactions on Information
  Theory}, vol.~51, no.~4, pp. 1568--1570, April 2005.

\bibitem{Fuchs98-detection}
J.-J. Fuchs, ``Detection and estimation of superimposed signals,'' in
  \emph{Proceedings of the 1998 {IEEE} International Conference on Acoustics,
  Speech, and Signal Processing {(ICASSP} '98)}, vol.~3, 12-15 May 1998, pp.
  1649 -- 1652 vol.3.

\bibitem{GribonvalNielsen03-spie}
R.~Gribonval and M.~Nielsen, ``Approximation with highly redundant
  dictionaries,'' in \emph{Wavelets: Applications in Signal and Image
  Processing, Proc. {SPIE'03}, San Diego, {USA}}, August 2003, pp. 216--227.

\bibitem{GribonvalVandergheynst04-exponential}
R.~Gribonval and P.~Vandergheynst, ``On the exponential convergence of matching
  pursuits in quasi-incoherent dictionaries,'' IRISA, Rennes, France, Tech.
  Rep. 1619, April 2004.

\bibitem{Donoho04-neighborly}
D.~L. Donoho, ``Neighborly polytopes and sparse solutions of underdetermined
  linear equations,'' Statistics Department, Stanford University, Tech. Rep.,
  December 2004.

\bibitem{Donoho05-high-dimensional}
------, ``High-dimensional centrosymmetric polytopes with neighborliness
  proportional to dimension,'' Statistics Department, Stanford University,
  Tech. Rep., January 2005.

\bibitem{McMullenShephard68-diagrams}
P.~McMullen and G.~C. Shephard, ``Diagrams for centrally symmetric polytopes,''
  \emph{Mathematika}, vol.~15, pp. 123--138, 1968.

\bibitem{Grunbaum03-convex}
B.~Gr\"unbaum, \emph{Convex Polytopes}, 2nd~ed., ser. Graduate Texts in
  Mathematics 221.\hskip 1em plus 0.5em minus 0.4em\relax New York:
  Springer-Verlag, 2003.

\bibitem{AvisFukuda92-pivoting}
D.~Avis and K.~Fukuda, ``A pivoting algorithm for convex hulls and vertex
  enumeration of arrangements and polyhedra,'' \emph{Discrete and Computational
  Geometry}, vol.~8, pp. 295--313, 1992.

\bibitem{Thie88-introduction}
P.~R. Thie, \emph{An Introduction to Linear Programming and Game Theory},
  2nd~ed.\hskip 1em plus 0.5em minus 0.4em\relax New York: John Wiley \& Sons,
  1988.

\bibitem{Schrijver98-theory}
A.~Schrijver, \emph{Theory of Linear and Integer Programming}.\hskip 1em plus
  0.5em minus 0.4em\relax Chichester, UK: John Wiley \& Sons Ltd, 1998.

\bibitem{LeeLewickiSejnowski00-ica-mixture}
T.-W. Lee, M.~S. Lewicki, and T.~J. Sejnowski, ``{ICA} mixture models for
  unsupervised classification of non-gaussian classes and automatic context
  switching in blind signal separation,'' \emph{IEEE Transactions on Pattern
  Analysis and Machine Intelligence}, vol.~22, no.~10, pp. 1078--1089, October
  2000.

\bibitem{MalioutovCetinWillsky04-optimal}
D.~M. Malioutov, M.~Cetin, and A.~Willsky, ``Optimal sparse representations in
  general overcomplete bases,'' in \emph{Proceedings of the {IEEE}
  International Conference on Acoustics, Speech, and Signal Processing
  {(ICASSP} '04)}, vol.~2, 17-21 May 2004, pp. II--793--796.

\bibitem{KvasnicaGriederBaotic04-mpt}
\BIBentryALTinterwordspacing
M.~Kvasnica, P.~Grieder, and M.~Baoti\'{c}, ``{Multi-Parametric Toolbox
  (MPT)},'' 2004. [Online]. Available: \url{http://control.ee.ethz.ch/~mpt/}
\BIBentrySTDinterwordspacing

\end{thebibliography}

\end{document}